%% file: main.tex
\documentclass[conference]{IEEEtran}
\IEEEoverridecommandlockouts
\usepackage{cite}
\usepackage{amsmath,amssymb,amsfonts}

\usepackage{algorithm}
\usepackage{algpseudocode}
\usepackage{array}
\usepackage[caption=false,font=footnotesize]{subfig}
\usepackage{graphicx}
\usepackage{textcomp}
\usepackage{stfloats}
\usepackage{url}
\usepackage{verbatim}
\usepackage{graphicx}
\usepackage{cite}
\usepackage{xcolor}
\usepackage[a4paper, total={184mm,239mm}]{geometry}
\usepackage{pifont}        
\usepackage{stackengine}   
\setstackEOL{\cr}          
\usepackage{threeparttable}
\usepackage{eso-pic}    
\usepackage[hidelinks]{hyperref} 
\usepackage{xcolor}

\newcolumntype{C}[1]{>{\centering\arraybackslash}m{#1}}

\graphicspath{{figs/}}
\def\BibTeX{{\rm B\kern-.05em{\sc i\kern-.025em b}\kern-.08em
    T\kern-.1667em\lower.7ex\hbox{E}\kern-.125emX}}
\begin{document}

\title{3D-ICE 4.0: Accurate and efficient thermal modeling for 2.5D/3D heterogeneous chiplet systems\\
}

\author{%
     \IEEEauthorblockN{ %
        Kai Zhu*, Darong Huang*, Luis Costeroº, David Atienza*%
    }
    \IEEEauthorblockA{%
        *\{kai.zhu, darong.huang, david.atienza\}@epfl.ch -- \textit{Embedded Systems Laboratory, EPFL. Switzerland}\\
        ºlcostero@ucm.es -- \textit{º Universidad Complutense de Madrid. Spain}%
    } %
}

\maketitle

\newcommand{\markblue}[1]{\textcolor{blue}{#1}}
\renewcommand{\markblue}[1]{#1}

\newcommand{\hook}{\ding{51}}   
\newcommand{\xmark}{\ding{55}}  

\newcommand{\halfhook}{%
  \stackengine{0pt}{\smash{\ding{51}}}{%
    \smash{\raisebox{0.05ex}{\scalebox{0.86}{\ding{55}}}}%
  }{O}{c}{F}{F}{S}%
}

\newcommand{\tparallel}[1][-10]{\ensuremath{\mathrel{\rotatebox[origin=c]{#1}{$\parallel$}}}}

\newcommand{\secstart}[1]{\label{p-#1-start}}
\newcommand{\secend}[1]{\label{p-#1-end}}
\newcommand{\secpages}[1]{pages \pageref{p-#1-start}--\pageref{p-#1-end}}

\newcommand{\opensrcmark}{\textsuperscript{\textasteriskcentered}}

\newcommand{\placecolbottomnoteL}[1]{%
  \AddToShipoutPictureFG*{%
    \AtTextLowerLeft{%
      \raisebox{-2em}[0pt][0pt]{
        \parbox{\columnwidth}{\raggedright\footnotesize #1}%
      }%
    }%
  }%
}

\begin{abstract}
The increasing power densities and intricate heat dissipation paths in advanced 2.5D/3D chiplet systems necessitate thermal modeling frameworks that deliver detailed thermal maps with high computational efficiency. Traditional compact thermal models (CTMs) often struggle to scale with the complexity and heterogeneity of modern architectures. This work introduces 3D-ICE 4.0\opensrcmark, designed for heterogeneous chip-based systems. Key innovations include: (i) preservation of material heterogeneity and anisotropy directly from industrial layouts, integrated with OpenMP and SuperLU MT-based parallel solvers for scalable performance, (ii) adaptive vertical layer partitioning to accurately model vertical heat conduction, and (iii) temperature-aware non-uniform grid generation. The results with different benchmarks demonstrate that 3D-ICE 4.0 achieves speedups ranging from \(3.61\times\)–\(6.46\times\) over state-of-the-art tools, while reducing grid complexity by more than \(23.3\%\) without compromising accuracy. Compared to the commercial software COMSOL, 3D-ICE 4.0 effectively captures both lateral and vertical heat flows, validating its precision and robustness. These advances demonstrate that 3D-ICE 4.0 is an efficient solution for thermal modeling in emerging heterogeneous 2.5D/3D integrated systems.
\end{abstract}

\begin{IEEEkeywords}
Thermal modeling, 2.5D/3D integration, chiplet systems, adaptive algorithm, non-uniform, parallel acceleration
\end{IEEEkeywords}

\input{s1_introduction}

\input{s2_related_work}

\input{s3_3d_ice_framework}

\input{s4_experiment}

\input{s5_conclusion}

\section*{Acknowledgment}

\markblue{This work was supported in part by the Swiss State Secretariat for Education, Research, and Innovation (SERI) through the SwissChips research project.}

\bibliographystyle{IEEEtran}
\bibliography{biblio}

\vfill

\end{document}

%% file: s1_introduction.tex
\section{Introduction}

\placecolbottomnoteL{\textasteriskcentered\ Open-source repository: https://github.com/esl-epfl/3d-ice.}


The explosive growth of compute-intensive workloads, from high-performance computing (HPC) to large language models (LLMs), has pushed chip power densities to unprecedented levels. Modern AI and HPC accelerators routinely dissipate hundreds of watts per package~\cite{latif2025single}, and multi-kilowatt servers concentrate substantial heat fluxes that, if not modeled and appropriately managed, can accelerate wear-out or trigger thermal throttling, ultimately degrading throughput~\cite{yi2023studying, huang2024evaluation}. Temperature has therefore become a first-order design constraint on par with performance and power~\cite{jiang2022fast}, requiring thermal models that can accurately predict full-chip temperature distributions under realistic and time-varying workloads.

Advanced 2.5D/3D heterogeneous integration (e.g., chiplet-based system-in-package or High Bandwidth Memory (HBM) stacked on logic dies) offers a promising way to scale compute and bandwidth beyond monolithic designs. However, these packaging approaches introduce complex heat flow paths~\cite{kim2022thermal}, lateral and vertical thermal coupling across heterogeneous materials (such as active silicon, interposer, underfill, micro-bumps, and through-silicon via (TSV) arrays), creating multi-scale paths with significant anisotropic characteristics~\cite{wang2023multiscale}. 
Hotspots in one die can increase the temperatures in neighboring dies. For example, in a tightly integrated 2.5D Graphics Processing Unit (GPU) system, the heat from a high-power die can raise the temperature of a neighboring HBM stack enough to induce a four-fold increase in the refresh rate~\cite{karfakis2025optimizing}, significantly reducing the throughput of the system. These realities require more detailed and accurate thermal models spatially and temporally to capture localized hotspots and thermal crosstalk in such complex architectures~\cite{jiang2022fast, yuan2021pact}.

Thermal analysis faces a non-trivial trade-off between accuracy and efficiency. Finite element methods (FEM) can accurately capture detailed heat conduction but are computationally and memory intensive, often requiring prohibitive costs for large-scale designs~\cite{pfromm2024mfit}. Compact thermal models (CTMs) based on equivalent thermal RC network (e.g., Hotspot~\cite{han20222}, 3D-ICE~\cite{terraneo20213d}, PACT~\cite{yuan2021pact}) run orders of magnitude faster than FEM and are widely used for thermal modeling and design-space exploration~\cite{wang2024atplace2}. 
However, this speed often comes at the cost of fidelity. Current CTMs have three common limitations: 1) reliance on coarse grids and homogenized materials and the difficulty of ingesting detailed, anisotropic geometries from industrial layouts (e.g., GDSII) into RC networks-which can misestimate on-die temperature gradients and hotspot magnitudes; 2) the use of a layer-based model with a division of one layer for each functional component, which can be insufficient to capture accurate cross-layer heat flow; and 3) the use of global uniform grids for simplicity, which either under-resolves hotspot regions or over-resolves other regions, yielding a suboptimal accuracy–efficiency trade-off as die size or layer count increase. Although 3D-ICE 3.1 supports non-uniform grids~\cite{huang2024evaluation}, a thermal-aware grid size selection strategy remains unexplored, which means users need to assign grid sizes to each region manually based on domain knowledge, which can be suboptimal for complex floorplans.
Therefore, a fine-grained CTM that preserves detailed geometry and material-aware lateral and vertical conduction in 2.5D/3D heterogeneous systems while remaining computationally efficient is needed. 

Based on these challenges and requirements, we present 3D-ICE 4.0, an open-sourced\opensrcmark\ framework and the key contributions include:

\begin{itemize}
  \item \textbf{Fine-detail layout and anisotropy-aware modeling:} An interface that automatically builds detailed models from industrial layout files, retaining detailed material distributions and anisotropic properties to enable accurate prediction of localized hotspots and cross-die interactions. Meanwhile, OpenMP~\cite{ayguade2008design}-based thermal matrix construction and a multi-threaded solver (SuperLU MT~\cite{li2011superlu}) are used for rapid steady-state and transient thermal simulation for large-scale chip models.
  \item \textbf{Adaptive layer division:} A strategy that subdivides layers with high vertical thermal resistance to better resolve cross-layer heat flow, improving accuracy for multilayer 2.5D/3D chiplet systems. 
  \item \textbf{Temperature-aware non-uniform grids:} A local grid refinement method that concentrates resolution in thermal critical regions to capture fine-grained heat flow variations, which can achieve a target error with far fewer grids than uniform meshes, improving simulation efficiency. 

\end{itemize}

%% file: s2_related_work.tex
\section{Related work}

Thermal modeling for 2.5D/3D chiplet systems spans three main families~\cite{atienza202520}: (i) analytical/Green's function methods, (ii) numerical methods (e.g., FEM and CTMs), and (iii) Machine Learning (ML) methods. 
Green's function approaches precompute the system's impulse responses and obtain temperature via convolution between Green's function and power maps, enabling rapid evaluations for fixed stacks~\cite{sultan2020fast}. However, changes in geometry, materials, or boundary conditions typically require recomputation of Green's function.
FEM, as implemented in commercial software such as Ansys~\cite{ansys_fluent_ug_2025R2} and COMSOL~\cite{comsol_ref_manual_63}, can offer high precision for complex structures and is often treated as reference solutions, but their runtime and memory costs limit rapid design space exploration and runtime thermal management. 
CTMs~\cite{han20222,terraneo20213d} approximate heat flow with equivalent thermal RC networks and are widely used for early design~\cite{wang2024atplace2} and thermal management~\cite{huang2024evaluation} due to orders-of-magnitude speedups over FEM~\cite{pfromm2024mfit}.
Recent works also explore ML surrogates for fast temperature prediction, such as convolutional neural networks (CNNs)~\cite{chhabria2021thermal}, graph neural networks (GNNs)~\cite{chen2022fast}, and physics-informed neural networks (PINNs)~\cite{liu2023deepoheat}. ML methods typically require extensive training data generated by FEM or CTMs, and struggle to generalize to unseen architectures~\cite{chen2022fast}. Consequently, CTMs remain a practical and reliable choice for thermal analysis of chiplet systems. 

\begin{table*}[htb]
\centering
\begin{threeparttable}
\caption{Comparison of representative compact thermal models.}
\label{tab:model_comparison}
\footnotesize
\begin{tabular}{C{2.0cm} | C{1.6cm} C{1.5cm} C{1.2cm} | C{2.0cm} C{2.0cm} C{1.5cm} | C{1.5cm}}
\hline
& \multicolumn{3}{c|}{\textbf{Geometry construction}} & \multicolumn{3}{c|}{\textbf{Grid generation}} & \multicolumn{1}{c}{\textbf{Solution}} \\[1pt]
\cline{2-4}\cline{5-7}\cline{8-8}
\textbf{Thermal models} & \textbf{Heterogeneous} & \textbf{Anisotropic} & \textbf{Automatic} & \textbf{Adaptive layer division} & \textbf{Non-uniform grids} & \textbf{Thermal-aware grids} & \textbf{Parallel acceleration}\\
\hline
Hotspot 7.0~\cite{han20222} & \hook &  &  &  &  &  & \\
3D-ICE 3.1~\cite{huang2024evaluation} &  &  &  &  & \hook &  & \\
PACT~\cite{yuan2021pact} & \hook &  &  &  &  &  & \hook \\
MFIT~\cite{pfromm2024mfit} & \hook & \hook &  &  & \hook &  & \hook \\
ARTSim 2.0~\cite{safari2025thermal} & \hook &  &  &  & \hook &  & \hook \\
\textbf{3D-ICE 4.0} & \hook & \hook & \hook & \hook & \hook & \hook & \hook \\
\hline
\end{tabular}
\end{threeparttable}
\end{table*}

Hotspot~\cite{han20222} pioneered RC-network-based CTMs for early-stage thermal analysis. To improve flexibility and efficiency, 3D-ICE 3.1~\cite{terraneo20213d} and ARTSim 2.0~\cite{safari2025thermal} introduce non-uniform grids.
As simulated systems become increasingly complicated, PACT~\cite{yuan2021pact} integrates a parallel SPICE solver to accelerate fine-grained thermal simulations. Similarly, MFIT~\cite{pfromm2024mfit} and ARTSim 2.0~\cite{safari2025thermal} use sparse linear algebra BLAS libraries and Python libraries SciPy and NumPy to speed up sparse solves, respectively. Despite these advancements, prior CTMs generally fall short in: 1) fast modeling of fine-detail heterogeneous layout, 2) accurate modeling of vertical heatflows, and 3) clear policies for non-uniform grids. The comparison of different representative CTMs is summarized in Table~\ref{tab:model_comparison}.



%% file: s3_3d_ice_framework.tex
\section{3D-ICE 4.0 framework}

\markblue{The typical CTM workflow (e.g., 3D-ICE 3.1) is shown in Fig.~\ref{s3_workflow_3dice}, and the new or updated stages in 3D-ICE 4.0 are highlighted in orange. After parsing the inputs (a), various components and a multilayer stack are built (b.1 and b.2). The stack is discretized into grids to obtain an RC network (b.3), which is transformed into a sparse linear system for steady-state or transient analysis (c and d). 
Unlike prior CTMs, 3D-ICE 4.0 supports heterogeneous and anisotropic materials, and directly imports detailed material distributions from industrial layout files (b.1.1). To better capture complex cross-layer heat flow in 2.5D/3D chiplet systems, we introduce an adaptive layer division strategy (b.2.1). In addition, a thermal-aware local grid refinement (b.3.1) is employed to achieve a better trade-off between accuracy and efficiency.}

Profiling shows that RC matrix assembly and sparse linear solutions dominate the runtime for CTMs and limit the scalability of detailed large-scale models. We address this bottleneck with OpenMP-based parallelized assembly~\cite{ayguade2008design} and a high-performance sparse solver SuperLU MT~\cite{li2011superlu}.
We will detail the novelties and functionalities of 3D-ICE 4.0 in the following subsections.


\begingroup
\setlength{\intextsep}{6pt}      
\setlength{\textfloatsep}{6pt}   
\begin{figure}[!htb]
    \centering
    \includegraphics[width=0.45\textwidth]{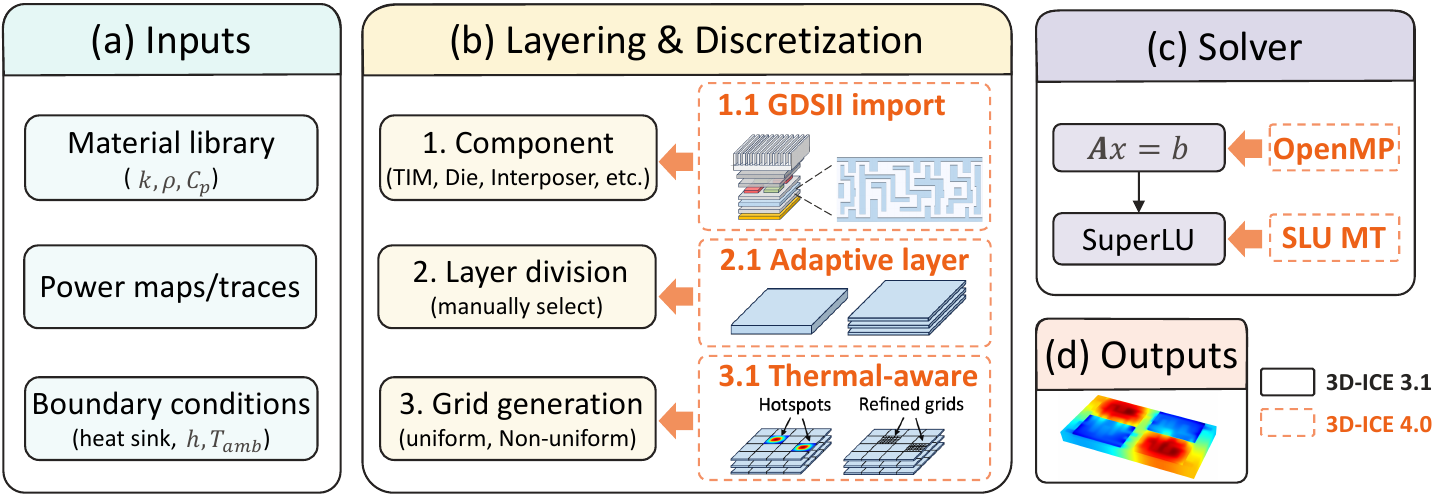}
    \vspace{-0.5em} 
    \caption{\markblue{Workflow of compact thermal modeling framework. Newly added/updated stages in 3D-ICE 4.0 are highlighted in orange.}}
    \label{s3_workflow_3dice}
\end{figure}
\endgroup

\subsection{Model generation with detailed layouts}

\begin{figure}[htb]
    \centering
    \includegraphics[width=0.47\textwidth]{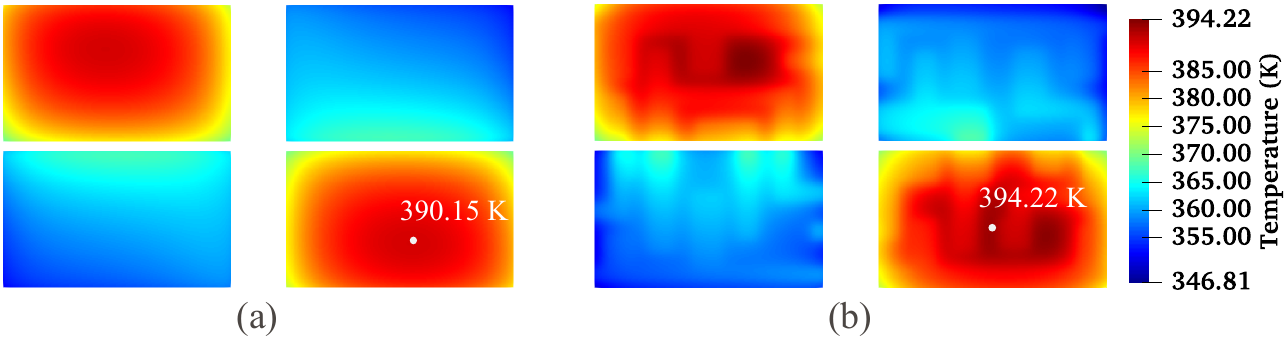}
    \vspace{-1em}
    \caption{Thermal map for a 4-chiplet system with (a) abstract homogeneous model, (b) heterogeneous model with detailed material distribution.}
    \label{s3_GDS_example}
\end{figure}

Accurate thermal simulation requires models that preserve the spatial heterogeneity of materials and layouts. In many prior CTMs~\cite{han20222,terraneo20213d}, complex layers are approximated by an effective medium calculated from the volume fractions of the constituent materials. To quantify the impact of this simplification, we model a 4-chiplet system in 3D-ICE 3.1 using homogenized layers, and the predicted hotspot is 390.15 K, as shown in Fig.~\ref{s3_GDS_example} (a). In contrast, when the detailed distribution of the material is retained, the hotspot increases to 394.22 K, as presented in Fig.~\ref{s3_GDS_example} (b). This discrepancy underscores the importance of modeling detailed layouts.


The manual construction of an RC network that preserves fine-detail material distributions is laborious and error-prone. We automate this step by directly importing material layouts from industry design files (e.g., GDSII files).
Two challenges exist: i) CTMs typically assume rectangular tiles to describe the floorplan, whereas real designs contain irregular geometries; and ii) GDSII layers often include overlapping objects spanning disparate length scales.
Our framework addresses these issues by i) incorporating equivalent models for irregular shapes~\cite{nie2022equivalent,wang2023multiscale}; and ii) applying an adaptive and layout-aware tiling scheme.
The extraction pipeline partitions each layer into non-overlapping tiles, computes each tile's intersection area with the polygons and paths in the GDSII layout, and then generates anisotropic per-tile equivalent models for subsequent thermal simulation. 

A quadtree-based tiling strategy (Algorithm 1) automates tile generation from layout geometry. The inputs include the GDS file, target layer, maximum recursion depth ($i_{max}$), and a stopping condition ($i_{con}$). 
All layer primitives (e.g., polygons, paths/FlexPaths) are first polygonized and merged into disjoint shapes via geometric union (line 2), then indexed with an R-tree built on their bounding boxes to accelerate spatial queries. The initial domain is the bounding box of the merged shapes (line 4), and we invoke the recursive \textsc{Subdivide} function. At each quadtree node, we query the R-tree for candidate shapes that intersect the current region (line 8), compute the overlap ratio $\rho$, and evaluate the stopping rule (line 10). If the criterion is satisfied, one tile record including position, size, and $\rho$, is emitted to the tile list; otherwise, the region is subdivided into four quadrants and processed recursively (lines 13-15). The fidelity can be adjusted flexibly by $i_{max}$ and $i_{con}$. 
For a complex layout in Fig.~\ref{s3_GDS_mesh_example}~(a), the resulting tile partition and associated $\rho$ map are shown in Fig.~\ref{s3_GDS_mesh_example} (b) and (c), which can be used to generate the detailed heterogeneous and anisotropic model for the layer.
The accuracy of the built model is validated in Section~\ref{experiment2}.

To accelerate simulations with detailed material distributions, 3D-ICE 4.0 parallelizes RC matrix assembly with OpenMP~\cite{ayguade2008design} and uses SuperLU MT~\cite{li2011superlu} for multi-thread preordering, factorization, and solution. As shown in Section~\ref{experiment1}, our framework can achieve a higher parallel efficiency than the SPICE-based solver in PACT~\cite{yuan2021pact}.


\begin{algorithm}[t]
\caption{Tile generation with quadtree-based subdivision}
\label{alg:tiles}
\textbf{Inputs:} \emph{gds}, \emph{layer}, $i_{\max}$, $i_{\text{con}}$ \\
\textbf{Output:} \texttt{tiles} \emph{(list of (position, size, $\rho$))}
\begin{algorithmic}[1]
  \State $S \gets \textsc{ReadGDS}(gds, layer)$
  \State $M \gets \textsc{Merge}(S)$
  \State $I \gets \textsc{BuildIndex}(M)$ \Comment{R-tree over merged polygons}
  \State $R \gets \textsc{MakeWindow}(M)$ \Comment{bounding region}
  \State \texttt{tiles} $\gets$ \textsc{Subdivide}$(I, R, i, i_{\max}, i_{\text{con}})$
  \State \Return \texttt{tiles}

  \Statex \dotfill  
  \Function{Subdivide}{$I, R, i, i_{\max}, i_{\text{con}}$}
    \State $C \gets \textsc{Candidates}(I, R)$ \Comment{query R-tree}
    \State $A_{\text{ov}} \gets \textsc{OverlapArea}(R, C)$, $\rho \gets A_{\text{ov}} / \textsc{Area}(R)$
    \If{Satisfy ($i_{\text{con}}$) \textbf{ or } {$i \ge i_{\max}$}}
       \State \texttt{tiles} $\gets$ $\,(R, \rho)\,$
    \Else
       \State $(R_1,R_2,R_3,R_4) \gets \textsc{SplitQuad}(R)$
       \For{$r \in \{R_1,R_2,R_3,R_4\}$} 
          \State \textsc{Subdivide}$(I, r, i{+}1, i_{\max}, i_{\text{con}})$
       \EndFor
    \EndIf
  \EndFunction
\end{algorithmic}
\end{algorithm}

\begingroup
\setlength{\intextsep}{6pt}      
\setlength{\textfloatsep}{6pt}   
\begin{figure}[htb]
    \centering
    \includegraphics[width=0.47\textwidth]{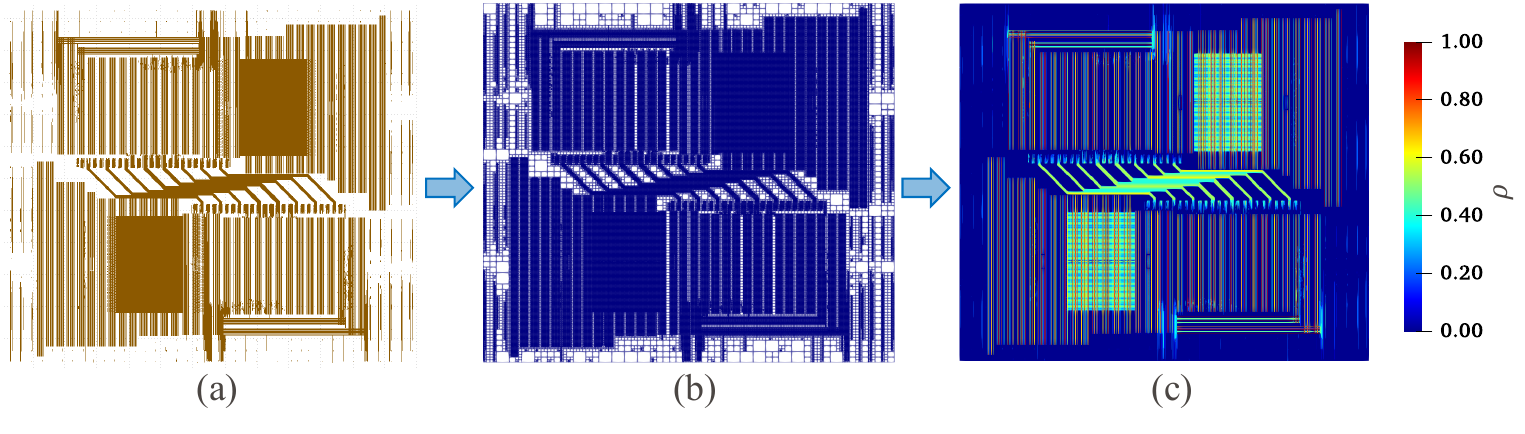}
    \vspace{-1em}
    \caption{Extraction of detailed material distribution, (a) GDSII layout of one layer, (b) generated tiles, (c) overlapped ratio for each tile.}
    \label{s3_GDS_mesh_example}
\end{figure}
\endgroup

\subsection{Adaptive layer division} \label{layer_division}
Vertical heat flow in 2.5D/3D heterogeneous chiplet systems is highly nonuniform and sensitive to material and geometric details~\cite{korouglu2023high}.
Conventional CTMs often align simulation layers (RC network layers) one-to-one with functional layers. This practice can induce large contrasts in vertical thermal resistances, where layers with high thermal resistance accumulate disproportionate temperature gradients and can not reflect the accurate inter-layer temperature profile~\cite{hankin2021hotgauge}.

To alleviate this problem, we introduce an adaptive layer division strategy during model construction, based on vertical thermal resistance. For layer $j$ with $N_j$ floorplan elements, the vertical resistance of element $i$ is \(R^{\perp}_{j,i}=h_j/(k^{\perp}_{j,i}\cdot A_{j,i})\), where $h_j$ is the layer thickness, $k^{\perp}_{j,i}$ is the equivalent vertical thermal conductivity, and $A_{j,i}$ is the area. The layer's equivalent vertical resistance is the parallel combination of each element \( R^{\perp}_{j}=\bigl(\sum_{i=1}^{N_j} \tfrac{1}{R^{\perp}_{j,i}}\bigr)^{-1} \). 
We quantify heterogeneity across an $n$-layer stack by the variance
\begin{equation}
\mathrm{Var}\!\left(R^{\perp}\right)
= \frac{1}{n}\sum_{j=1}^{n}\left(R^{\perp}_j - \bar{R}^{\perp}\right)^{2}
\label{eq:variance_resistance}
\end{equation}
where $\bar{R}^{\perp}$ is the mean vertical resistance.

Layers that dominate $\mathrm{Var}\!\left(R^{\perp}\right)$ or exhibit large $R^{\perp}_j$ are iteratively subdivided into multiple simulation sublayers, reducing and redistributing the per-layer vertical resistance. Iteration stops when the variance falls below a prescribed threshold or reaches a maximum depth of iteration. 
\markblue{The refinement is performed before the simulation using the material and geometric parameters, and the associated overhead is negligible.} The accuracy gains of the layer division strategy are validated in Section~\ref{experiment2}.  

\begingroup
\setlength{\intextsep}{3pt}      
\setlength{\textfloatsep}{3pt}   
\begin{figure}[tb]
    \centering
    \includegraphics[width=0.45\textwidth]{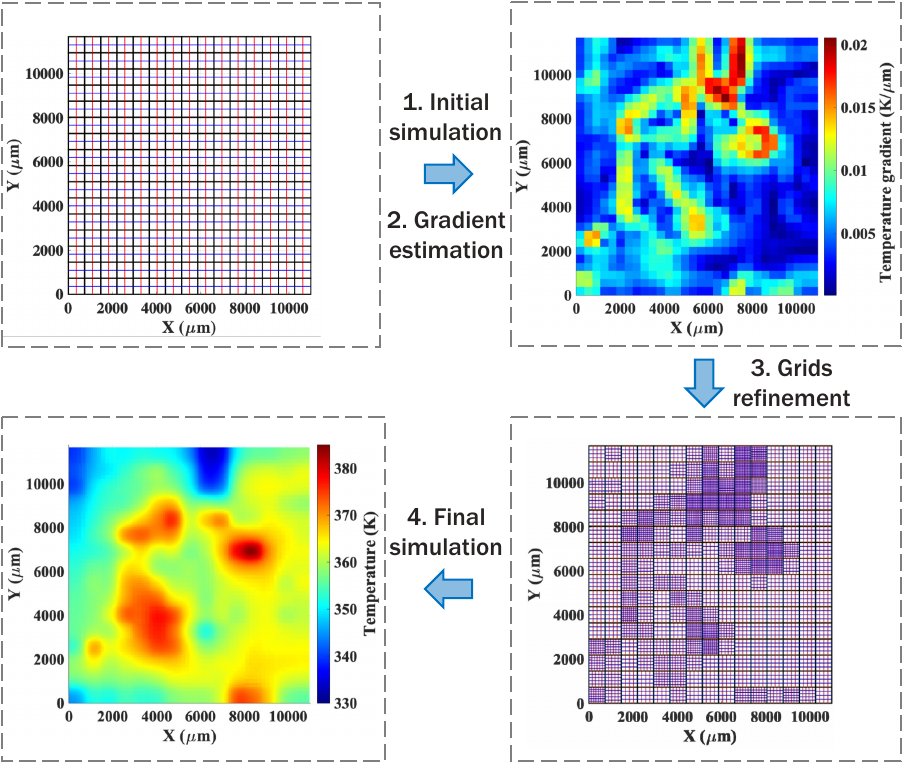}
    \vspace{-0.5em}
    \caption{Workflow of the generation of thermal-aware non-uniform grids and simulation.}
    \label{non_uniform_mesh}
\end{figure}
\endgroup

\subsection{Local grid refinement} \label{localgrids}

3D-ICE 3.1~\cite{huang2024evaluation} and ARTSim 2.0~\cite{safari2025thermal} support non-uniform grids within one layer and across different layers, offering more flexibility and a better accuracy-efficiency trade-off than uniform grids. Finer grids can be used in specific regions of interest without modifying the grids in other regions to reduce computational overhead and simulation time. 
However, a systematic policy for choosing the local grid granularity has not been fully explored. Therefore, we propose an automatic refinement strategy guided by local temperature gradients, as shown in Fig.~\ref{non_uniform_mesh}. 
Starting from a coarse mesh, we perform an initial solve and estimate the average temperature gradient $G$ for each floorplan element. Elements exhibiting large gradients (e.g., near hotspots) are selectively refined to better capture steep variations. For a given element, the grid size in the x-direction is chosen as:
\begin{equation}
\mathrm{Gridsize}_{x} = l_x \cdot \alpha \cdot \left(\frac{G_{base}}{G+\epsilon}\right)
\label{eq:gridsize_x}
\end{equation}%
where $l_x$ is the length of the floorplan element in the x-direction, $G_{base}$ and $\alpha$ are control parameters for the refinement of the grid and $\epsilon > 0$ avoids division by zero. 
The calculated $\mathrm{Gridsize}_{x}$ needs to be regulated to $[l_{min},l_x]$, where $l_{min}$ is the minimum granularity permitted by runtime and memory constraints. The size of the grid in the y-direction can be computed in a similar way. The efficiency gains of this temperature-aware local refinement over uniform grids are demonstrated in Section~\ref{experiment3}.

%% file: s4_experiment.tex
\begin{figure}[tb]
    \centering
    \includegraphics[width=0.45\textwidth]{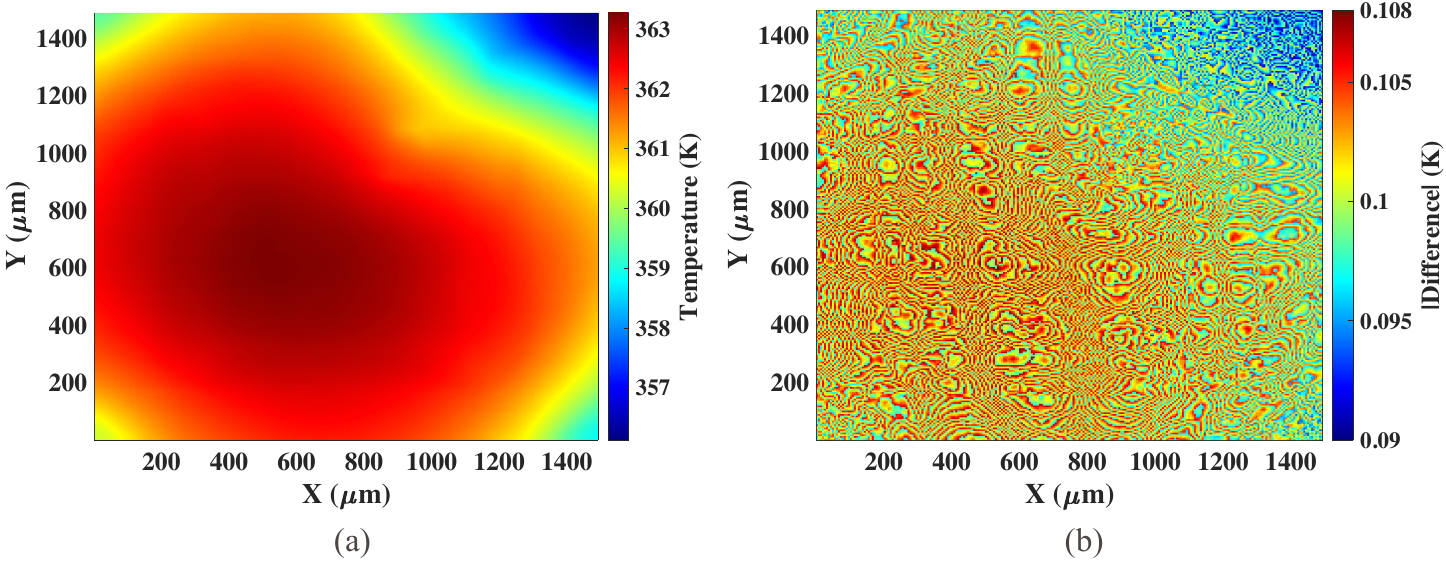}
    \vspace{-1em}
    \caption{(a) Temperature map from 3D-ICE 4.0 with a uniform \(256\times256\) grid, (b) the absolute difference map between 3D-ICE 4.0 and PACT.}
    \label{case1_pact}
\end{figure}

\section{Experimental results}

In this section, three validation examples are provided to evaluate the accuracy and efficiency improvement of 3D-ICE 4.0. All simulations run on a server with two AMD EPYC 9734 CPUs and 1.0-TB DDR5 memory, under Ubuntu 24.04. 

\subsection{Case1: Accuracy and efficiency validation} \label{experiment1}

To validate the accuracy of our framework, we reproduce the example of the state-of-the-art parallel compact thermal model PACT~\cite{yuan2021pact}, using a uniform \(256\times256\) grid. Fig. \ref{case1_pact} (a) shows the steady-state temperature map produced by 3D-ICE 4.0, and Fig. \ref{case1_pact} (b) shows the absolute difference with respect to PACT. 
The small discrepancy demonstrates the accuracy of 3D-ICE 4.0 with respect to PACT, which has been validated against Hotspot~\cite{han20222} and COMSOL~\cite{yuan2021pact}, and also provides indirect validation of 3D-ICE 4.0 against these simulators.

\begin{figure}[htb]
    \centering
    \includegraphics[width=0.48\textwidth]{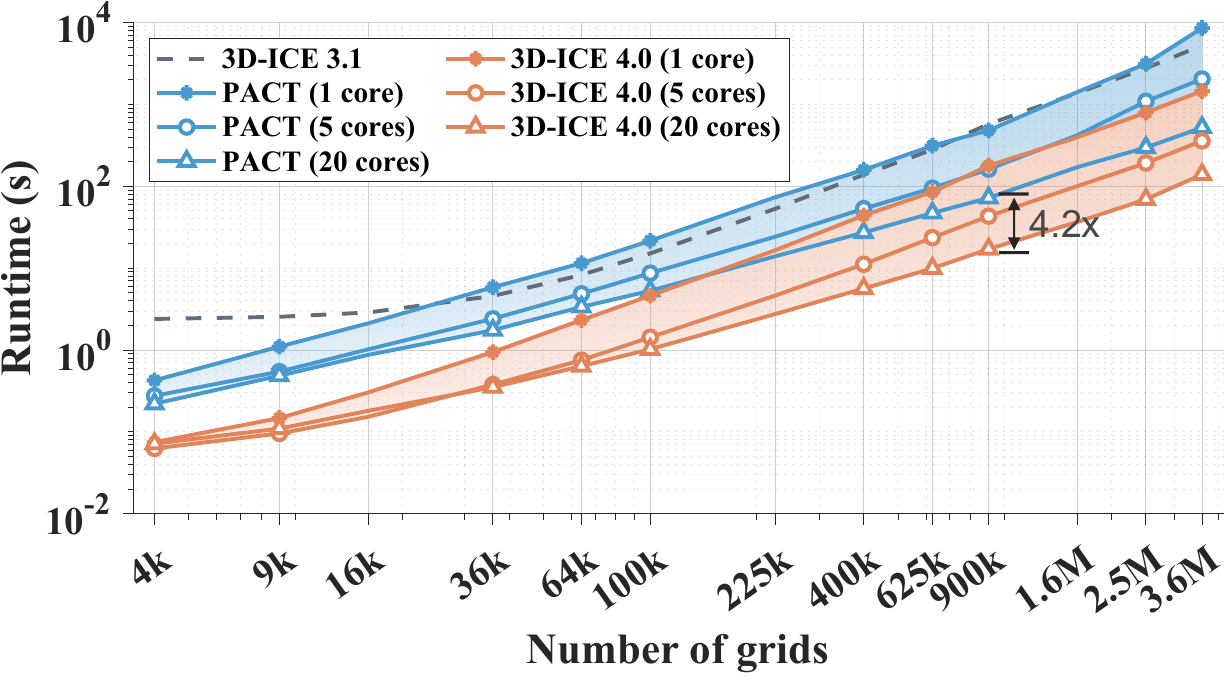}
    \vspace{-1em}
    \caption{Runtime for 3D-ICE 3.1, PACT and 3D-ICE 4.0 with 1, 5, and 20 cores versus different problem sizes.}
    \label{case1_grids}
\end{figure}

\begin{figure}[htb]
    \centering
    \includegraphics[width=0.48\textwidth]{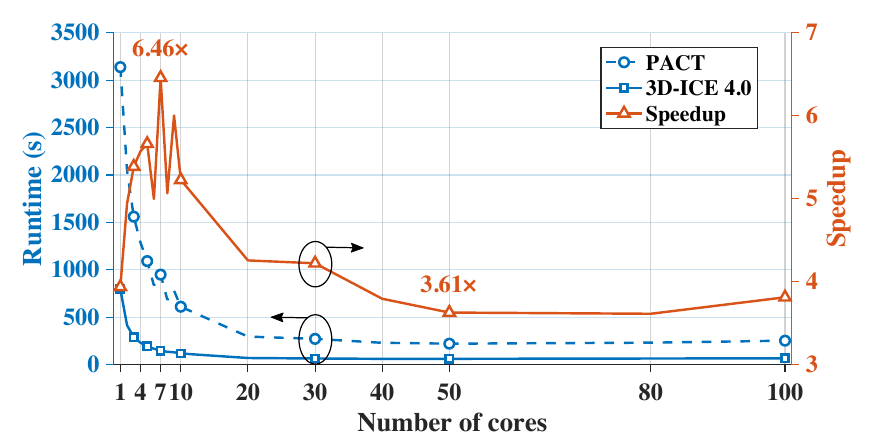}
    \vspace{-1em}
    \caption{Runtime for PACT and 3D-ICE 4.0 with different numbers of cores.}
    \label{case1_cores}
\end{figure}

To assess parallel efficiency, we sweep the grid resolution and compare the simulation time for serial solver 3D-ICE 3.1, PACT, and 3D-ICE 4.0 using 1, 5, and 20 CPU cores in Fig. \ref{case1_grids}. As problem size increases, both PACT and 3D-ICE 4.0 exhibit higher speedup ratios, and for any fixed core count, 3D-ICE 4.0 consistently achieves shorter runtimes than PACT across the test, indicating better parallel efficiency and scalability. 

\begingroup
\setlength{\intextsep}{4pt}      
\setlength{\textfloatsep}{4pt}   
\begin{figure*}[htb]
  \centering
  \includegraphics[width=\textwidth]{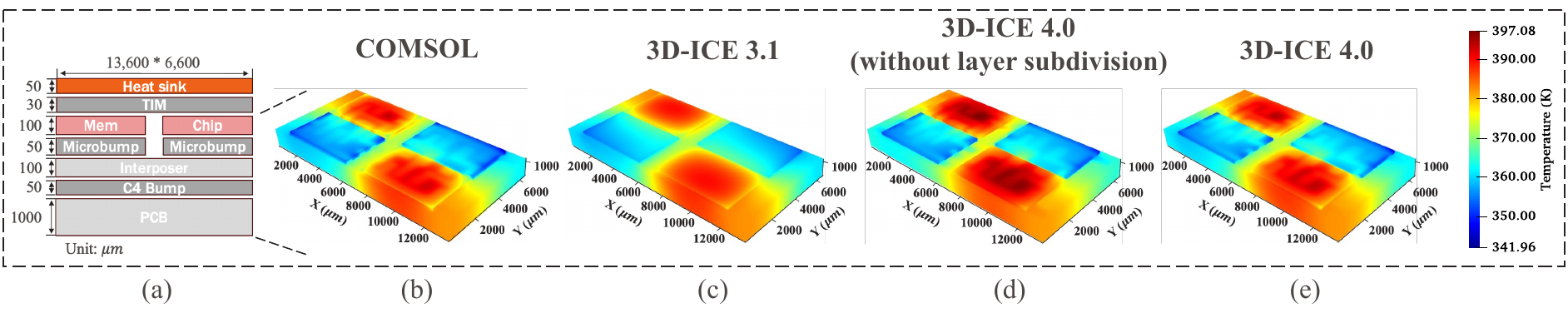}
  \vspace{-2em}
  \caption{Simulated geometry and 3D temperature distribution. (a) Geometry and dimension of the chiplet system, temperature map from (b) COMSOL, (c) 3D-ICE 3.1, (d) 3D-ICE 4.0 without layer subdivision, (e) 3D-ICE 4.0 with layer division.}
  \label{Case2_geometry}
\end{figure*}
\endgroup

To compare the speedup ratio with more different core numbers, we fix the grid count as 2.5 million, then vary the core count and compare 3D-ICE 4.0 with PACT, as shown in Fig. \ref{case1_cores}. 
It can be found that the runtime keeps decreasing with increasing cores up to around 30, after which gains saturate. Across all core counts tested, 3D-ICE 4.0 is \(3.61\times\)–\(6.46\times\) faster than PACT. 
For time division analysis, when the number of cores is set to 7, PACT requires 949.0 s to finish the simulation, 148.2 s for netlist import and matrix assembly, and 800.8 s for solving. 
In contrast, 3D-ICE 4.0 finishes in 146.9 s, with only 0.8 s for parsing the inputs and assembling the matrix system, and 146.1 s for solving. 
In addition, the peak memory usage of our framework (35.28~GB) is 27.3\% lower than that of PACT (48.43~GB), which highlights the memory efficiency of our approach.

\begingroup
\setlength{\intextsep}{2pt}      
\setlength{\textfloatsep}{2pt}   
\begin{table}[htb]
\centering
\begin{threeparttable}
\caption{Thermal properties of different components}
\label{tab:thermal_params}
\footnotesize
\setlength{\tabcolsep}{3pt}
\renewcommand{\arraystretch}{1.05}
\begin{tabular}{@{} C{1.5cm} C{1.9cm} C{1.5cm} C{1.5cm} @{}}
\hline
\textbf{Component} &
\textbf{Thermal conductivity [\(\mathrm{W/(m\cdot K)}\)]} &
\textbf{Density  [\(\mathrm{kg/m^{3}}\)]} &
\textbf{Heat capacity [\(\mathrm{J/(kg\cdot K)}\)]} \\
\hline
Heatsink               & 385   & 8,900  & 387   \\
TIM                    & 5     & 2,500  & 1,000 \\
Chip, PCB              & 130   & 2,300  & 700   \\
microbump, C4 bump     & \(5.5 (\tparallel), 113(\perp)\)  & 7,380 & 250 \\
Underfill              & 1.5   & 1,400  & 1,100   \\      

\hline
\end{tabular}
\end{threeparttable}
\end{table}
\endgroup

\subsection{Case2: Modeling with detailed material distribution and accuracy gains by adaptive layer division} \label{experiment2}

After initial validation, we evaluate how detailed material layouts and adaptive layer division affect accuracy of thermal modeling of a 2.5D chiplet system~\cite{scheffler2025occamy}. The simulated system is shown in Fig. \ref{Case2_geometry} (a), material properties are listed in Table~\ref{tab:thermal_params}, and the interposer layout is imported from the GDSII file in \cite{hammoud2023openfasoc}.
The steady-state temperature map from COMSOL serves as the reference (Fig. \ref{Case2_geometry} (b)). Using 3D-ICE 3.1 with a homogenized model fails to capture the temperature gradients, as shown in Fig. \ref{Case2_geometry} (c). 
Incorporating detailed material distributions in 3D-ICE 4.0 can capture these temperature gradients, but still deviates from COMSOL due to the limited vertical resolution, as shown in Fig. \ref{Case2_geometry} (d).  
After 8 iterations of adaptive layer division, as illustrated in Section~\ref{layer_division}, results from 3D-ICE 4.0 align closely with COMSOL, as presented in Fig. \ref{Case2_geometry} (e), underscoring the importance of layer division. 
For a fair comparison, we match the number of grid unknowns in 3D-ICE 3.1 and 3D-ICE 4.0 to the degrees of freedom (DoFs) of the COMSOL setup. 

\begingroup
\setlength{\intextsep}{5pt}      
\setlength{\textfloatsep}{5pt}   
\begin{figure}[htb]
  \centering
  \includegraphics[width=0.45\textwidth]{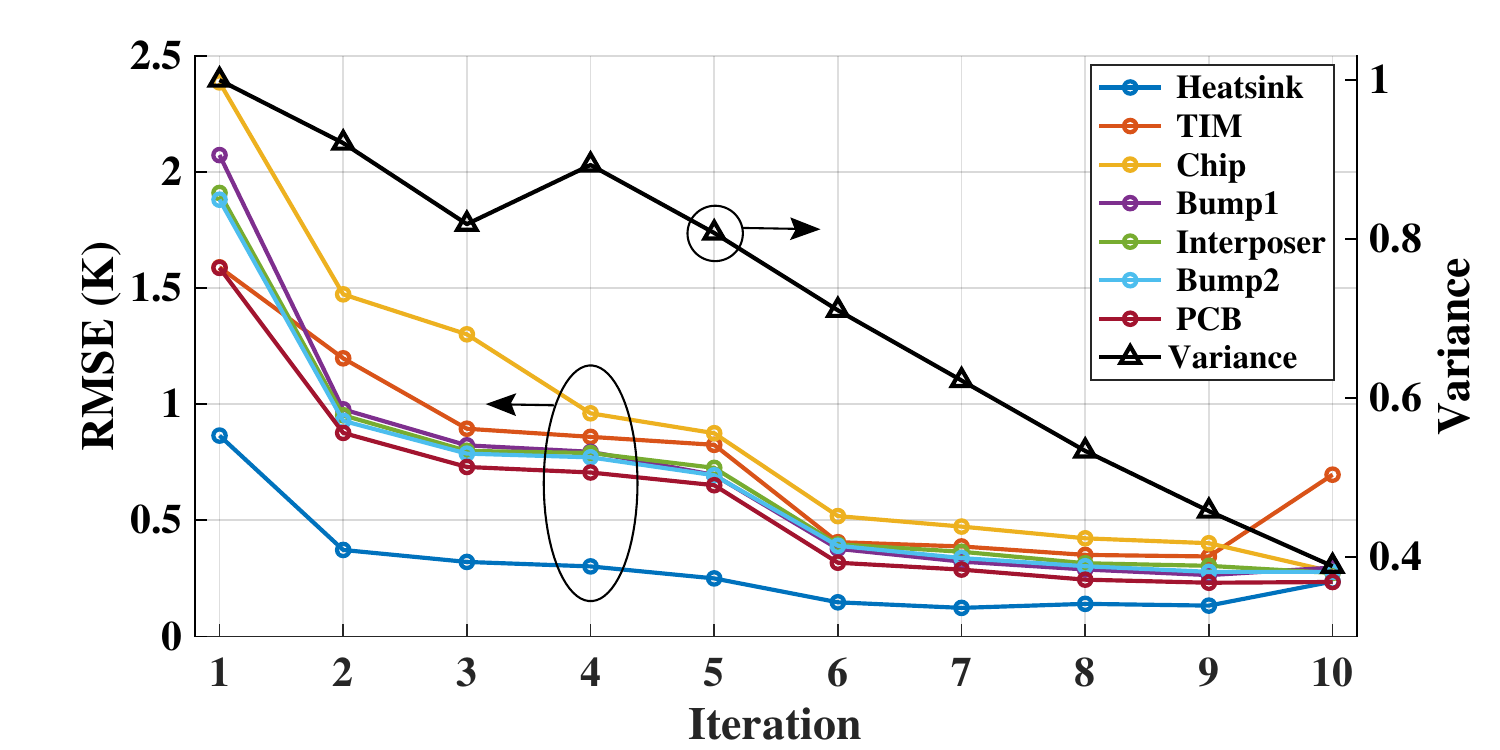}
  \vspace{-1em}
  \caption{RMSE for functional layers during layer division iteration.}
  \label{case2_error_iteration}
\end{figure}
\endgroup

\begin{figure*}[htb]
  \centering
  \includegraphics[width=\textwidth]{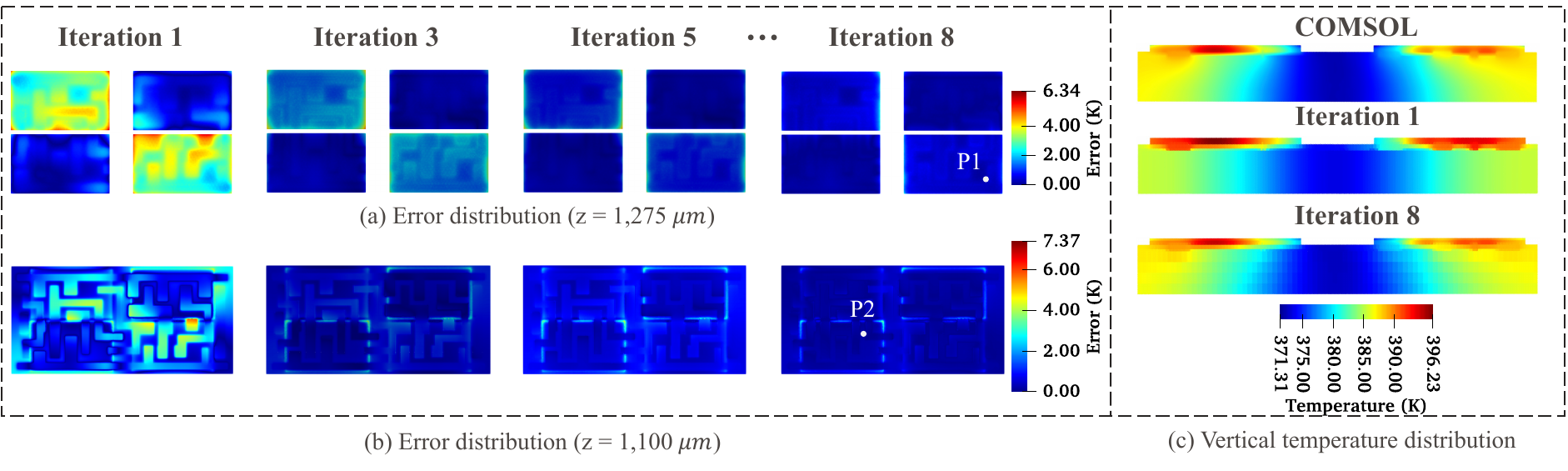}
  \vspace{-2em}
  \caption{Cross-sectional error and temperature distribution during iteration. (a) error distribution of chip ($z = 1,275 \, \mu\mathrm{m}$) and (b) interposer ($z = 1,100 \, \mu\mathrm{m}$) from 3D-ICE 4.0 compared with COMSOL, (c) vertical temperature slice at the center point along the $z$-axis.}
  \label{Case2_slice}
\end{figure*}

Across iterations, the layers with the larger vertical thermal resistance are adaptively split into thinner layers with the lower vertical resistance. The normalized variance of vertical resistance across all simulation layers (Fig. \ref{case2_error_iteration}) decreases as the distribution becomes more uniform, and the root mean square error (RMSE) for each functional layer with respect to COMSOL falls accordingly (Fig. \ref{case2_error_iteration}). 
For the chip and interposer, Fig. \ref{Case2_slice} (a) and (b) compare the results from 3D-ICE 4.0 with COMSOL: the first iteration shows notable errors, especially in high-power regions, whereas after 8 iterations-by which time the TIM and PCB are each subdivided into four sub-layers, and the chip into two-the agreement is markedly improved. 
It is worth noting that we keep the total number of grids per functional layer constant across iterations, which means the accuracy improvement arises from improved vertical temperature modeling rather than an increased grid count. 
A temperature slice through the stack's center along the $z$-axis is shown in Fig. \ref{Case2_slice} (c), which further confirms that the vertical temperature profile converges toward the COMSOL reference after iteration. 

\setlength{\dbltextfloatsep}{8pt} 
\setlength{\abovecaptionskip}{4pt}
\setlength{\belowcaptionskip}{0pt}

To assess the accuracy of our framework for transient thermal analysis, we apply Gaussian-modulated sinusoidal power inputs:
 \begin{equation}
 P(t) = P_0 + P_0 \exp\!\left(-\frac{(t - t_0)^2}{\tau^2}\right)\,\sin(\omega t)
 \label{eq:Power}
 \end{equation}
 where $\tau^2 = 0.1$, $\omega = 10\pi$, $t_0 = 0.5$, $P_0 = 25W$ and $50W$ for the memory and logic dies, respectively. 
 We run a transient simulation and record temperatures at two observation points, the locations are shown in Fig. \ref{Case2_slice}.
 Fig. \ref{case2_transient} compares transient temperature traces of 3D-ICE 4.0 without layer division, with layer division, and the COMSOL reference. Without layer division, the temperature traces deviate markedly from COMSOL, while the traces with layer division agree well with COMSOL in both amplitude and timing, underscoring the necessity of layer division and validating that 3D-ICE 4.0 can accurately capture fast, complex thermal dynamics.

\begingroup
\setlength{\intextsep}{2pt}      
\setlength{\textfloatsep}{2pt}   
\begin{figure}[htb]
    \centering
    \includegraphics[width=0.45\textwidth]{figs/s4_Case2_transient}
    \vspace{-1em}
    \caption{Transient temperature at two observation points: 3D-ICE 4.0 without/with adaptive layer division versus COMSOL, and absolute error with respect to COMSOL, (a) point 1, (b) point 2.}
    \label{case2_transient}
\end{figure}
\endgroup

\subsection{Case3: Efficiency improvement by local refined grids} \label{experiment3}

After the accuracy validation of adaptive layer division, we evaluate the efficiency gains of local refined grids introduced in Section~\ref{localgrids}. The test case is a 100 $\mu\mathrm{m}$-thick die with a heat sink attached on top; the die power map is taken from the gate-level simulation in \cite{yuan2021pact}. Following the workflow in Fig.~\ref{non_uniform_mesh}, we start from a coarse mesh (960 grids), perform an initial simulation, compute the spatial distribution of temperature gradients, and refine grids according to the local gradients. 

We set $\alpha = 1$ in Eq.~\eqref{eq:gridsize_x} and sweep $G_{base}$ to control the total grid count, then run a final simulation with the local refined grids. The RMSE versus grid count is shown in Fig.~\ref{case3_1layer_compare}. For comparison, we also solve with uniform grids over a similar range of grid counts and plot the error curve. 
The non-uniform strategy reaches a target error with fewer grids: to achieve RMSE \(<0.3\,\mathrm{K}\), the uniform strategy requires 11{,}760 grids versus 8{,}714 for the non-uniform strategy (25.9\% fewer). For RMSE \(<0.25\,\mathrm{K}\), the requirements are 15{,}360 and 11{,}787 cells, respectively (23.3\% reduction). 

The corresponding absolute error maps for the 15{,}360 uniform grids and the 11{,}787 non-uniform grids are also shown in Fig. \ref{case3_1layer_compare}. The peak error for uniform grid simulation can be up to 1 K, and concentrates in regions with high temperature gradients. By using finer grids in these regions, non-uniform grid simulation reduces the peak error to 0.6 K. Therefore, local-refined grids not only improve efficiency (fewer grids for a target RMSE) but also enhance the ability to capture steep temperature gradients.

\begingroup
\setlength{\intextsep}{2pt}      
\setlength{\textfloatsep}{2pt}   
\begin{figure}[htb]
    \centering
    \includegraphics[width=0.45\textwidth]{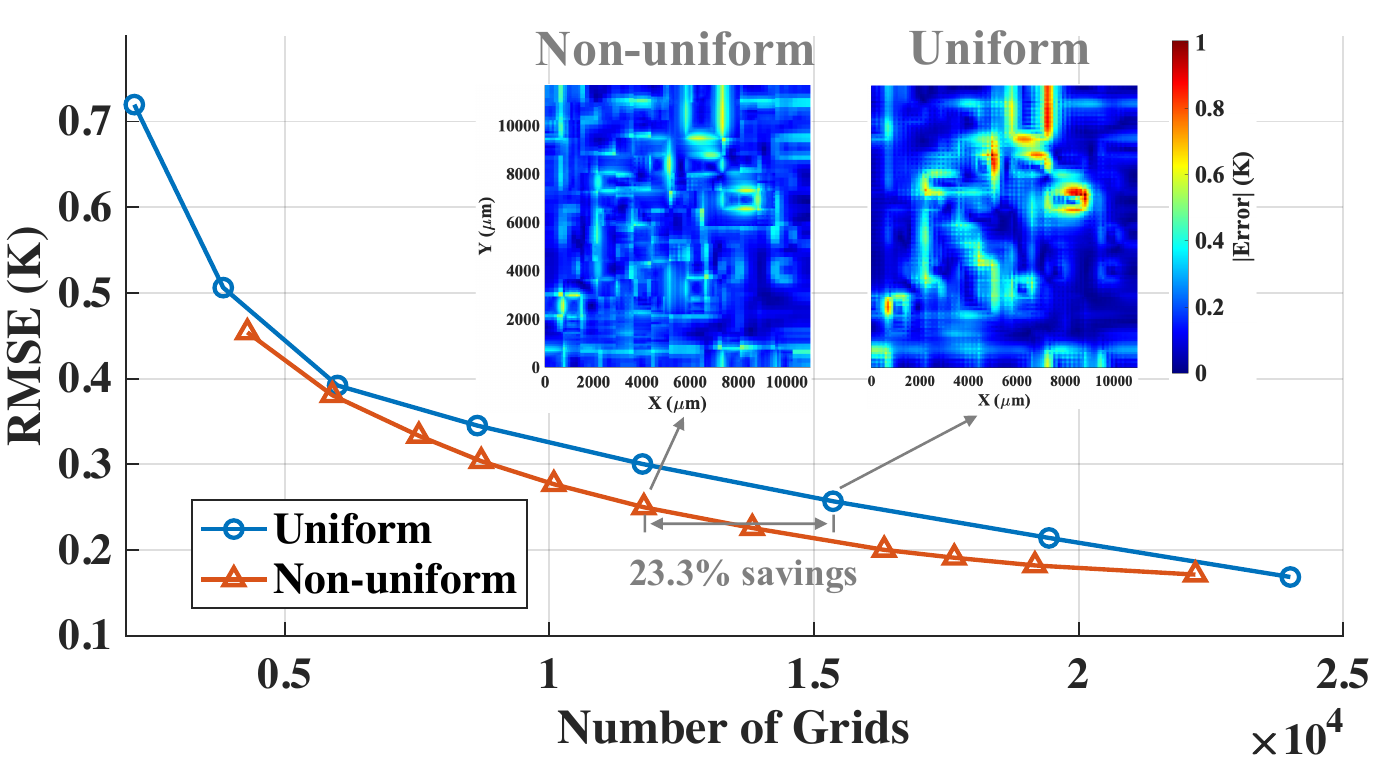}
    \vspace{-1em}
    \caption{Error with different numbers of grids for uniform grids and non-uniform grids.}
    \label{case3_1layer_compare}
\end{figure}
\endgroup

%% file: s5_conclusion.tex
\section{Conclusion}
In this work, we have presented 3D-ICE 4.0, a new compact thermal modeling framework, which preserves material heterogeneity and anisotropy from industrial layouts, introduces adaptive layer partitioning, applies temperature-aware non-uniform grids to reduce complexity, and employs parallel acceleration for scalable thermal simulation. Experiments show that 3D-ICE 4.0 can achieve \(3.61\times\)–\(6.46\times\) speedups over PACT and closely matches COMSOL in both steady-state and transient simulations. For a target error, the grid strategy can reduce unknowns by \(23–26\%\), improving efficiency without sacrificing accuracy. 
These results indicate that 3D-ICE 4.0 is promising for faster design-space exploration and more reliable runtime thermal management for emerging 2.5D/3D chiplet systems.